\documentclass[fleqn,usenatbib]{mnras}

\usepackage{mathptmx}
\usepackage{txfonts}
\usepackage{lipsum}  

\usepackage[T1]{fontenc}

\DeclareRobustCommand{\VAN}[3]{#2}
\let\VANthebibliography\thebibliography
\def\thebibliography{\DeclareRobustCommand{\VAN}[3]{##3}\VANthebibliography}

\usepackage{graphicx}	
\usepackage{orcidlink}
\usepackage{times}
\usepackage{xspace}
\usepackage{comment}
\usepackage{listings}
\usepackage{nth}

\newcommand{\codename}[1]{\textcolor{black}{\sc #1}\xspace}        
\newcommand{\simulationname}[1]{\textcolor{black}{\sc #1}\xspace}  
\newcommand{\techjargon}[1]{\textcolor{black}{\tt #1}\xspace}      

\newcommand{\swift}{\codename{Swift}}

\newcommand{\flamingo}{\simulationname{Flamingo}}
\newcommand{\cowl}{\simulationname{Cosmo-OWLS}}
\newcommand{\owl}{\simulationname{OWLS}}
\newcommand{\bahamas}{\simulationname{Bahamas}}
\newcommand{\antilles}{\simulationname{Antilles}}

\newcommand{\lcdm}{$\Lambda$CDM\xspace}

\title[Baryonic matter power spectrum suppression in \flamingo]{The \flamingo project: Baryon effects on the matter power spectrum}

\author[M. Schaller et al.]{
Matthieu Schaller\,\textsuperscript{\orcidlink{0000-0002-2395-4902}}$^{\,1,2}$\thanks{E-mail: \url{mschaller@lorentz.leidenuniv.nl}},
Joop Schaye\textsuperscript{\orcidlink{0000-0002-0668-5560}}$^{2}$,
Roi Kugel\textsuperscript{\orcidlink{0000-0003-0862-8639}}$^{2}$,
Jeger C. Broxterman\textsuperscript{\orcidlink{0000-0002-8155-5977}}$^{1,2}$
\& Marcel P. van Daalen\textsuperscript{\orcidlink{0000-0002-8801-4911}}$^{2}$

\\
$^{1}$Lorentz Institute for Theoretical Physics, Leiden University, PO Box 9506, NL-2300 RA Leiden, The Netherlands\\
$^{2}$Leiden Observatory, Leiden University, PO Box 9513, NL-2300 RA Leiden, The Netherlands
}

\date{Accepted XXX. Received YYY; in original form ZZZ}

\pubyear{\the\year{}}

\begin{document}
\label{firstpage}
\pagerange{\pageref{firstpage}--\pageref{lastpage}}
\maketitle

\begin{abstract}
The effect of baryon physics associated with galaxy formation onto the
large-scale matter distribution of the Universe is a key uncertainty in the
theoretical modelling required for the interpretation of Stage IV cosmology
surveys. We use the \flamingo suite of simulations to study the baryon response
due to galaxy formation of the total matter power spectrum. We find that it is
only well converged for simulation volumes in excess of $200^3~{\rm Mpc}^3$. We
report results for simulations of varying feedback intensity, which either match
the X-ray inferred gas fractions in clusters and the $z=0$ stellar mass
function, or shifted versions of the data, as well as for different
implementations of AGN feedback. We package our results in the form of a
Gaussian process emulator which can rapidly reproduce all the simulations'
predictions to better than one per cent up to the comoving wavenumber $k =
10~h\cdot {\rm Mpc}^{-1}$ and up to $z=3$ for all the feedback models present in
the \flamingo suite. We find that the response becomes stronger, the range of
scales affected increases, and the position of the minimum of the response moves
to smaller scales as the redshift decreases. We find that lower gas fractions in
groups and clusters lead to a stronger response and that the use of collimated
jets instead of thermally driven winds for AGN feedback enhances the
effect. Lowering the stellar masses at fixed cluster gas fractions also
increases the magnitude of the response. We find only a small (one per cent at
$k<10~h\cdot {\rm Mpc}^{-1}$) dependence of our results on the background
cosmology, but a wider range of cosmology variations will be needed to confirm
this result. The response we obtain for our strongest feedback models is
compatible with some of the recent analyses combining weak lensing with external
data. Such a response is, however, in strong tension with the X-ray inferred gas
fractions in clusters used to calibrate the \flamingo model.
\end{abstract}



\section{Introduction}
\label{sec:introduction}

Over nearly three decades, our standard model of cosmology, the \lcdm model, has
received substantial scrutiny from the community and successfully passed a
multitude of stress tests \citep[see e.g.][]{Dodelson2020, Lahav2022}. This vast
program, designed to find faults in the model, understand its limitations, and
identify possible extensions is continuing in this decade with exceedingly
demanding precision tests, generally grouped under the ``Stage IV cosmology
probe'' label. Many of these tests are focusing on the growth of the large scale
structure (LSS) and are providing independent constraints from the geometric
probes, such as the baryon acoustic oscillations (BAO) or Type 1a supernovae, or
the analysis of fluctuations of the cosmic microwave background (CMB). Most of
these programs were designed to shed some light on the nature of both dark matter
and dark energy as well as to explore some of the tensions currently emerging
between orthogonal probes \citep[see e.g.][]{Abdalla2022}.

The many different LSS tests (e.g. cosmic shear, galaxy clustering,
redshift-space distortions, CMB lensing, Sunyaev–Zel'dovich power spectra, and
combinations thereof) probe the matter content of the Universe and its
distribution across many different length-scales and at multiple epochs
throughout cosmic time. As the scales probed become smaller with each generation
of instruments, the challenge of making accurate theoretical predictions
grows. Many of the probes mentioned above are now exploiting information well
into the non-linear regime, where perturbation theory is not sufficient
anymore. The main approach used over the last twenty years has been to resort to
the ``halo model'' analytic formalism \citep[e.g.][]{Seljak2000, halofit,
  Asgari2023} itself (usually) calibrated on the results of $N$-body simulations
\citep[e.g.][]{Takashi2012, Mead2016}. More recently, the ability to run a
sufficiently large number of cosmological simulations has allowed an alternative
approach based on the direct interpolation between simulations (typically via
emulators) to predict many quantities required for data analysis
\citep[e.g.][]{Heitmann2016, Lawrence2017, DeRose2019, Euclid2019, Bocquet2020,
  Angulo2021, StoreyFisher2024}.

Both the halo model and emulators trained on pure $N$-body simulations would
suffice if the scales probed were not affected by the behaviour of
baryons. Whilst on large scales, $k \lesssim 0.1~h\cdot{\rm Mpc}^{-1}$, the
joint baryon and dark matter fluid behaves similarly to a pure dark matter model
(with initial conditions accounting for BAO), studies based on simulations
including hydrodynamics and galaxy formation effects have shown that the matter
field on smaller scales deviates significantly from the pure $N$-body
predictions \citep[e.g.][]{VD2011, VD2020, Schneider2015, Mummery2017,
  Springel2018, Salcido2023, Pakmor2023, Schaye2023, Wang2024} and that
neglecting this effect will result in catastrophic systematic errors
\citep[e.g.][]{Semboloni2011}. The amplitude of the baryonic effect depends on
uncertain feedback processes and is thus difficult to predict. However, the
amplitude is understood to depend on observables such as the baryonic content of
groups and clusters \citep[e.g.][]{Semboloni2011, Semboloni2013, VD2015,
  Chisari2019, VD2020, Mead2020, Salcido2023, vanLoon2024} and halo models
reproducing these observables confirm the importance of baryon effects
\citep{Debackere2020}.  The modelling of these deviations, commonly referred to
as ``baryon effects'', is crucial for the interpretation of future surveys
probing the matter distribution deep into the non-linear regime.

Thanks to its speed and flexibility, the community's preferred approach thus
far, has been to use relatively simple analytic prescriptions to correct the
predictions of $N$-body simulations \citep[e.g.][]{Schneider2015, Arico2021} or
halo models \citep[e.g.][]{Mead2021}. These correction procedures themselves
come with free parameters which could, in principle, be marginalised over when
inferring cosmological information from survey data \citep[e.g.][]{Asgari2021,
  Arico2023,Bigwood2024}. Whilst being powerful, these extended halo models and
other ``baryonification'' procedures also come with some drawbacks. The most
important ones being the relatively simple nature of the models and the implicit
assumption that the baryon physics is independent of the chosen background
cosmology (however, see \citealt{Arico2021} for a model going beyond this
assumption). These models also do not come with a clear ab-initio prediction for
the strength of the correction required, though attempts have been made to link
some of their parameters to observables to restrict the range of valid input
parameters \citep[e.g.][]{Schneider2015, Mead2020, Debackere2020, Schneider2022,
  Troster2022, Arico2023, Ferreira2024}.

An alternative approach is to exploit hydrodynamical simulations. Their cost is
much larger than the evaluation of corrective methods, but recent advances have
allowed for large suites of simulations that vary the input parameters of their
sub-grid models to be run \citep[e.g.][]{LeBrun2014, BAHAMAS, CAMELS,
  Salcido2023, Schaye2023}. Among the advantages of this approach are the
self-consistent nature of the modelling and the relative ease with which the
simulated data can be connected to observables.

Whilst knowing and understanding the exact feedback mechanisms in galaxies would
provide an ab-initio prediction for the properties of groups and clusters and
thus of the baryon effect on the matter power spectrum, this is far beyond our
current understanding of galaxy formation. Numerical simulations thus come with
sub-grid recipes which are calibrated to match specific observables. The natural
choice, based on the discussion above, is to target the gas or baryon fractions
in groups and clusters of galaxies. The \bahamas \citep{BAHAMAS},
\simulationname{Fable} \citep{Henden2018}, and \antilles \citep{Salcido2023}
projects all took this approach to set the value of their free
parameters. Thanks to the rather large simulation volumes probed, the effects of
galaxy formation on the matter power spectrum from the \cowl \citep{LeBrun2014}
and \bahamas simulations are often used as references for the range of possible
outcomes \citep[e.g.][]{Amon2022,Preston2023}. They have also been used as input
to some of the simpler corrective models described above
\citep[e.g.][]{Mead2021, Arico2023}. Running large suites of simulations varying
parameters around the best-fitting model is possible, but a more efficient
approach to restrict the plausible parameter space is to build on the work of
\cite{VD2020}. By analysing virtually all simulations from the literature at the
time and building on the results from \cite{Semboloni2011, Semboloni2013}, they
formalised robustly the connection between group and cluster baryonic content
and the effect of baryons on the matter power spectrum. \cite{Salcido2023} built
on this idea to construct an emulator trained on the $400$ simulations of their
\antilles suite to predict the baryonic response of the matter power spectrum as
a function of various combinations of cluster properties. Their emulation
approach relating the unknown effect to some (almost) observable quantities
allows to use their tool in cosmology inference (thanks to its speed) whilst
providing meaningful data-driven priors.

In this study, we follow similar steps using the \flamingo suite of cosmological
simulations \citep{Schaye2023, Kugel2023}. These simulations cover much larger
simulated volumes than the \antilles suite and have been calibrated, using
modern machine-learning-based techniques, to reproduce the observed \emph{gas
fractions} in groups and clusters. Whilst the connection to the power spectrum
is less direct for the gas fraction, it is more easily observable than the total
baryon fraction and the gas fraction is typically much larger than the stellar
fraction for the relevant halo masses. The \flamingo simulations have been shown
to reproduce a series of observables of the cluster population
\citep{Schaye2023, Braspenning2023, Kay2024} and thus offer a good baseline for
the understanding of the effect of baryons on the matter power
spectrum. Furthermore, the use of variations of the base model, where the
observables have been shifted in a systematic fashion \citep{Kugel2023}, allows
for a direct connection between the baryonic response and the observables the
simulations were calibrated to. Additionally, the simulations themselves have
already been used to investigate the role of baryons on the so-called S8 tension
\citep{McCarthy2023, McCarthy2024, Elbers2024}. In this paper, we construct a
simple and fast Gaussian process emulator \citep[for an introduction,
  see][]{Rasmussen2006} predicting the baryon effects on the power spectrum as a
function of redshift and simulation calibration targets. By publicly releasing
our emulator, we provide a simple and efficient way of incorporating the results
of the \flamingo suite of simulations in analysis pipelines of upcoming surveys.

This paper is organised as follows. In Sec.~\ref{sec:simulations}, we introduce
the simulation model used and verify the convergence of the results with the
simulation volume. In Sec.~\ref{sec:emulator}, we describe the procedure used to
construct our baryon response emulator and validate it. In
Sec.~\ref{sec:results}, we explore some results obtained using the emulator.
Finally, we offer some conclusions in Sec.~\ref{sec:conclusions}.

\section{Simulations \& Power-spectra measurements}
\label{sec:simulations}

In this section, we present the simulations used for this study
(Sec.~\ref{ssec:flamingo}), describe how the matter power spectra are measured
(Sec.~ \ref{ssec:ps_measurements}) and analyse the convergence of the results
(Sec.~ \ref{ssec:convergence}).

\subsection{The \flamingo suite of simulations}
\label{ssec:flamingo}

The \flamingo simulations and the strategy used to calibrate their free
parameters to match relevant observables are described in \cite{Schaye2023} and
\cite{Kugel2023}. We provide here a brief summary of the key components.

The simulations were performed using the \swift simulation code
\citep{SWIFT}\footnote{Publicly available, including the exact version used for
the \flamingo simulations, at \url{www.swiftsim.com}.}, a fully open-source
coupled cosmology, gravity, hydrodynamics, neutrino and galaxy formation
code. The gravity is solved using a 4$^{\rm th}$-order fast-multipole-method
\citep[see e.g.][]{Cheng1999} coupled to a Particle-Mesh method in Fourier space
for the long-range interactions using the splitting method of
\cite{Bagla2003}. Cosmological neutrinos are evolved using the $\delta
f$-method of \cite{Elbers2021}. The equations of hydrodynamics are evolved using
the SPHENIX \citep{Borrow2022} flavour of Smoothed Particle Hydrodynamics.

The hydrodynamical simulations include subgrid prescriptions for radiative
cooling following \cite{Ploeckinger2020}, an entropy floor at high densities and
star formation using the method of \cite{Schaye2008}, stellar feedback using
kinetic winds \citep{DallaVecchia2008, Chaikin2022}, and the chemical enrichment
model of \cite{Wiersma2009}. Supermassive black holes and thermally driven AGN
feedback are modelled following \cite{Springel2005}, \cite{Booth2009} and
\cite{Bahe2022}. The models with ``jet'' AGN feedback (see below) alternatively
use the method of \cite{Husko2022} to produce feedback using collimated jets.

As is the case in all galaxy formation models, the simulations of the \flamingo
suite have free parameters in their subgrid recipes. The approach chosen for
this project to calibrate the feedback parameters was to construct a Gaussian
process emulator, trained on a Latin hypercube of simulations, to predict the
observables as a function of the subgrid parameters. This emulator was then used
to calibrate the models against observational data, as presented in detail by
\cite{Kugel2023}. The simulations were chosen to reproduce the $z=0$ galaxy
stellar mass function as well as the gas fractions in groups and clusters
inferred from X-ray and weak-lensing data. This choice is similar to the one
made for the \bahamas simulations \citep{BAHAMAS} but using a more systematic
approach via the emulator. 

Besides its added objectivity over calibration by hand, the use of an emulator
to set the free parameters of the model offers an additional advantage: the
possibility to rapidly generate simulated models where the data is shifted by
particular amounts with respect to observations. In the case of the cluster gas
fractions, we created different models where the observed gas fractions are
shifted up and down compared to the results by $\pm N\sigma$, where $\sigma$ is
the scatter in the data \citep[see][for the exact definitions]{Kugel2023}. Once
the emulator has been fitted to these shifted data points, we run a full
simulation using the predicted subgrid parameter values. These simulations are
labelled as ``fgas$\pm N\sigma$'' in Table \ref{tab:run_variations}. Similarly,
we generated models fitting shifted versions of the stellar mass function,
effectively lowering/increasing the mass of every galaxy by $\pm N\sigma$,
where, $\sigma$ is the systematic error on the measurements. These runs are
labelled as ``$M_*\pm N\sigma$'' in Table \ref{tab:run_variations}. Finally,
we ran two simulations using the jet model of AGN for two different shifts of
the gas fractions. These runs are labelled as ``Jets fgas$\pm N\sigma$'' in
Table \ref{tab:run_variations}.

Outside of the tests in Sec.~\ref{ssec:cosmo_dep}, the simulations used in this
study adopt as values of the cosmological parameters the maximum likelihood
values from the DES year 3 data release \citep{Abbott2022} combined with
external probes (their ``3$\times$2pt + All Ext.''  model: $\Omega_{\rm m}=0.306,~
\Omega_{\rm b}=0.0486,~\sigma_8=0.807,~h=0.681,~n_{\rm s}=0.967, \sum m_\nu c^2=~
0.06{\rm eV}$). The initial conditions (ICs) were generated using the
\codename{MonofonIC} code \citep{Hahn2021, Elbers2022} using a 3-fluid formalism
with a separate transfer function for each of the dark matter, gas, and
neutrinos. The ICs used partially fixed modes \citep{Angulo2016}, setting the
amplitudes of modes with $(kL)^2 < 1025$ to the mean variance, where $L$ is the
side-length of the simulated box and $k$ the modes' wave-number. The multi-fluid
approach \citep{Rampf2021} used by \codename{MonofonIC} allows to obtain a
perfect match of the gravity-only and full-hydrodynamics power spectrum on the
largest scales.

\begin{table}
\centering
\caption{The nine different simulations of the \flamingo suite used in this
  study. All models assume the same cosmology and were run in $1~{\rm Gpc}^3$
  volumes using the same initial conditions. The first column gives the
  simulation names used by \citet{Schaye2023}. The next two columns give the
  amount by which the gas fractions or stellar masses were shifted when
  calibrating the model. The last column indicates whether the model was run with
  the fiducial (thermal isotropic) AGN model or with the collimated jet
  implementation. The last three columns are the three parameters used as input
  to our baryon response emulator.}
\label{tab:run_variations}
\begin{tabular}{|| l | c | c | c||} 
 \hline
 Simulation name & fgas $\pm N\sigma$ & $M_*$~$\pm N\sigma$ & Jet fraction \\
 \hline\hline
 fgas$+2\sigma$      & +2 &  0 & 0\%   \\
 L1\_m9 (fgas$+0\sigma$)          &  0 &  0 & 0\%   \\
 fgas$-2\sigma$      & -2 &  0 & 0\%   \\
 fgas$-4\sigma$      & -4 &  0 & 0\%   \\
 fgas$-8\sigma$      & -8 &  0 & 0\%   \\
 $M_*-\sigma$        &  0 & -1 & 0\%   \\
 $M_*-\sigma$~fgas$-4\sigma$      & -4 & -1 & 0\%   \\
 Jets                &  0 &  0 & 100\% \\
 Jets fgas$-4\sigma$ & -4 &  0 & 100\% \\
 \hline
\end{tabular}
\end{table}

\subsection{Measurements of power-spectra and baryonic response}
\label{ssec:ps_measurements}

The measurement of power spectra is performed over the course of the simulation
at 122 different times between $z=30$ and $z=0$. At $z<3$, we obtain a power
spectrum measurement after every redshift interval $\Delta z=0.05$.

For each computation, we deposit the particles on a regular grid of size $256^3$
using a triangular-shaped-clouds approach which is self-consistently compensated
in Fourier space \citep[see e.g.][]{Jing2005}. Once the density field is
computed, we compute its power spectrum using a fast-Fourier transform.  To
reach $k$-modes beyond the spacing of the base grid, we use the foldings
technique of \cite{Jenkins1998} with a folding factor of $4$ between iterations
and use $7$ iterations. This allows us to measure the matter power spectrum up
to scales $k > 1000~h\cdot{\rm Mpc}^{-1}$, much beyond the needs of this
project.

We measure the total matter power spectrum (i.e. the sum of the gas, dark
matter, stars, black holes, and neutrinos) in the hydrodynamical simulations and
perform the same computation in the dark-matter-only simulations, where the
matter field is represented by a single type of particle. By using the same grid
and folding settings between all the runs, we obtain measurements in the same
bins in $k$-space and can thus simply obtain the matter power spectrum ratios,
i.e. the baryonic effect on the power spectrum, by dividing the two
spectra\footnote{The raw matter power-spectra for all the simulations of the
\flamingo suite have been made publicly available on the website of the project:
\url{https://flamingo.strw.leidenuniv.nl/}.}.

\begin{figure*}
  \includegraphics[width=\textwidth]{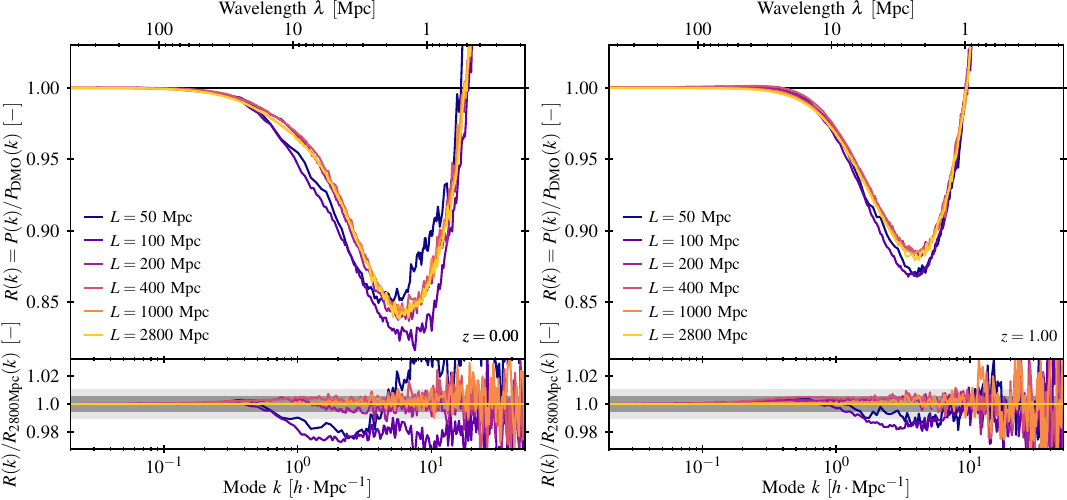}
\vspace{-0.5cm}
\caption{\textit{Top:} The baryonic response at redshift $z=0$ (left) and $z=1$
  (right) for the fiducial \flamingo model extracted from simulations with
  different volumes, labelled by the box side length. \textit{Bottom:} The ratio
  of the response in each simulation to the response obtained in the largest
  simulation ($L=2800~{\rm Mpc}$). The shaded regions correspond to fractional
  errors of $0.5$ and $1$ per cent respectively. At both redshifts, the results
  are converged at the $1$ per cent level for simulations with volumes in excess
  of $200^3~{\rm Mpc}^3$ up to $k\approx 10~h\cdot{\rm Mpc}^{-1}$. Simulations
  with a side-length of $400~{\rm Mpc}$ are not converged to better than $0.5$
  per cent even on scales $k\approx 1~h\cdot{\rm Mpc}^{-1}$, with a larger
  deviation at higher redshift.  Note also the reduction in the level of noise
  when larger simulation volumes are used.}
\label{fig:simulations:convergence}
\vspace{-0.3cm}
\end{figure*}

The $z=0$ baryonic responses for the simulations listed in
Table~\ref{tab:run_variations} were already presented by \citealt{Schaye2023}
(their Fig. 22) and compared to the results of the \bahamas \citep{BAHAMAS} and
{\sc Millennium-TNG} \citep{Pakmor2023} predictions. \cite{Schaye2023} also
showed that the results of the various \flamingo runs are in excellent agreement
with the model of \cite{VD2020} relating the baryonic response to the mean
baryon fraction in clusters of mass $M_{500{\rm c}}=10^{14}~{\rm M}_\odot$,
where $M_{500{\rm c}}$ is the mass within a radius enclosing a spherical
overdensity $500$ times larger than the critical density of the Universe.

We note that the simulations using the ``jet'' implementation of AGN feedback
displayed a response slightly smaller than unity ($0.9968$) on the largest
scales \citep[see the bottom right panel of Fig. 22 of][]{Schaye2023}. As the
response in the $k$-range affected by feedback is much larger than this
difference and since this small offset creates artifacts for the emulator
construction, we decided to renormalise the response of the jet models in what
follows.

\subsection{Convergence test with simulation volume}
\label{ssec:convergence}

As the effect of baryon physics on the matter power spectrum is thought to be
driven by haloes of different masses for the different $k$ modes \citep[see
  e.g.][]{Semboloni2011, VD2020, Debackere2020, Mead2020, Salcido2023,
  vanLoon2024}, it is important to ensure that the haloes responsible are well
sampled in the simulation volume used for the analysis and are not affected by
cosmic variance. To verify this, we present the baryonic response of the matter
power spectrum at $z=0$ and $z=1$ for different simulations in the \flamingo
suite using exactly the same galaxy formation model and at fixed resolution in
Fig.~\ref{fig:simulations:convergence}. We use the fiducial \flamingo model here
(i.e.  ``fgas$+0\sigma$, $M_*+0\sigma$, $0\%$ jets'', labelled as the
``L1\_m9'' model in Table \ref{tab:run_variations}) but we verified that the
results are similar for other models. The different line colours correspond to
simulations using cubic volumes whose side-lengths are indicated in the figure
and range from $50~{\rm Mpc}$ to $2.8~{\rm Gpc}$ \citep[the ``L2p8\_m9'' model
  of][]{Schaye2023}.

For volumes with side-length in excess of $200~{\rm Mpc}$ at $z=0$, we find that
the baryonic response is converged at the $2$ per cent level for the entire
range of $k$ values probed by current and upcoming surveys. A convergence to
better than $0.5$ per cent is only achieved for the simulation with a $1~{\rm
  Gpc}$ side-length.  At $z=1$ (right panel), the convergence at the $1$
per cent level is also achieved for volumes $>200^3~{\rm Mpc}^3$. However, at a
more precise level, the simulations are further from the converged result on
larger scales than at $z=0$. As smaller $k$ values are affected by haloes of
larger masses, this difference in convergence can be interpreted as the absence
of the rarer objects at higher redshift. In the analysis that follows, we will
make use of simulations with a volume of $1~{\rm Gpc}^3$, well within the regime
where the results are converged.

Based on this simulation volume analysis, we caution that matter power spectrum
responses extracted from simulations such as \simulationname{Illustris}
\citep{Vogelsberger2014}, \simulationname{Eagle} \citep{Schaye2015},
\simulationname{Horizon-AGN} \citep{Chisari2018}, \simulationname{Simba}
\citep{Dave2019}, and \simulationname{Camels} \citep{Delgado2023} --- all using
simulated volumes $\lesssim 100^3~{\rm Mpc}^3$ --- are likely not
converged. Furthermore, if the difference from a converged result in these
models has the same sign as we find in the \flamingo runs, one could expect the
results of these studies to have \emph{overestimated} the baryonic response of
these models at $k<4~h\cdot{\rm Mpc}^{-1}$. On the other hand, from our analysis
the response reported for the \simulationname{TNG} model \citep{Springel2018}
would be converged for their largest volume ($\approx 300^3~{\rm Mpc}^3$). This
is confirmed by the results of the \simulationname{Millennium-TNG} simulations,
using a very similar model in a $740^3~{\rm Mpc}^3$ volume, reported by
\cite{Pakmor2023}. We note that if the baryonic response is dominated by the
activity in haloes of lower mass than in \flamingo (as in
e.g. \simulationname{Eagle} or \simulationname{TNG}), the response might be
converged in smaller volumes already as the number density of the haloes
responsible will be less affected by cosmic variance than for our model.

The results of our convergence tests here are consistent with the findings of
\cite{VD2011} who found that the total matter power spectra at $z=0$ are not
converged in hydrodynamical simulations with volumes $< (100/h~{\rm Mpc})^3$
(their appendix A) and of \cite{VD2020} who found the response to be converged
only for volumes $\gtrsim (100/h~{\rm Mpc})^3$ (their Appendix A).

\section{Gaussian Process Emulator}
\label{sec:emulator}

In this section, we describe the procedure used to construct our power spectrum
response emulator (Sec.~\ref{ssec:construction}) and the validation steps we
performed to assess its quality (Sec.~\ref{ssec:validation}).

\subsection{Construction of the emulator}
\label{ssec:construction}

To construct our emulator, we make use of the nine runs listed in
Table~\ref{tab:run_variations} and measure the baryonic response as described in
Sec.~\ref{ssec:ps_measurements}. We measure the ratio $R(k) = P(k)/P_{\rm
  DMO}(k)$ for $31$ values of $k$ logarithmically spaced between $10^{-1.5}$ and
$10^{1.5}~h\cdot{\rm Mpc}^{-1}$ (i.e. $\Delta \log_{10}k = 0.1$) at seven
different redshifts ($z=0., 0.5, 1.0, 1.5, 2.0, 2.5$, and $3.0$). We thus have a
coarse representation of $R$ as a function of five input numbers: $k$, $z$, and
the three model parameters used to describe the baryonic physic in the
simulations (fgas$\pm N\sigma$, $M_*\pm N\sigma$, and jet fraction) as given in
Table \ref{tab:run_variations}. Note that even though our simulations are with
the fiducial AGN model \emph{or} the jet-based AGN model, we treat the AGN model
as a continuous number going from $0\%$ jet to $100\%$ jet. Similarly, we do not
restrict the emulator to only accept input parameter values within the range
where it was trained. Whilst these are somewhat uncertain interpolations and
extrapolations, we prefer to be able to cover a wide range of scenarios with our
model, even if some of them may not be reproduced by actual simulations.

We then train the Gaussian process emulator provided by the publicly
available \techjargon{python} package \codename{swiftemulator}
\footnote{\url{https://swiftemulator.readthedocs.io/}}
\citep{Kugel2022}, itself an overlay specialised in scaling relations extracted
from simulations of the commonly used
\codename{george}\footnote{\url{https://george.readthedocs.io/}}
package \citep{george}.

The emulator was trained on a small number of $k$ bins to reduce the amount of
internal data generated and speed up the prediction process. As we will show
below, the number of bins we used is sufficient to achieve better than $1$ per cent
relative accuracy for the range of input parameters relevant to our
application. We note that we just used the tools above as-is and that no
specific hyper-parameter tuning was necessary. The only hyper-parameter choice
made was to force the emulator to assume a mean model of $R(k)=1~\forall~k$ and
emulate the difference from this imposed model rather than letting the emulator
freely choose a mean $R(k)$ --- typically a polynomial --- around which to emulate
differences.

When using the emulator to make predictions, we use the quadruplet ($z$, ${\rm
  fgas\pm N\sigma}$, ${\rm M}_*\pm N\sigma$, jet fraction) to get the values of
$R$ at the $31$ points along the $k$-axis defined above. We then use spline
interpolation to compute $R$ at the exact $k$ values of interest. By predicting
the value of $R$ for all the $31$ $k$-bins at the same time, we can more rapidly
return $R(k)$ for a range of $k$ values at once, which is the most common
scenario. As we empirically have $R(k) = 1$ for all $k<10^{-1.5}~h\cdot{\rm
  Mpc}^{-1}$ in our models, we extend our emulator to simply return $R=1$ on all
scales larger than the training range. This effectively allows us to predict the
baryonic response in the range $k\in(-\infty, 10^{1.5}]~h\cdot{\rm Mpc}^{-1}$,
  meaning that our emulator can easily be coupled to other tools predicting
  $P_{\rm DMO}(k)$ to obtain a $P(k)$ including baryon effects over the whole
  range of scales relevant to current cosmological analysis needs. \\

We note that the entire prediction step described above takes approximatively
$1~{\rm ms}$ to be performed on a single compute core. This implies that our
emulator can be used in cosmology model inference searches (e.g. using an MCMC
sampler) without leading to problematic time overheads. It could also be used as
a simple extension to the commonly used Boltzmann solvers such as
\codename{CAMB} \citep{CAMB} or \codename{CLASS} \citep{CLASS} on top of their
already-implemented non-linear extensions such as \codename{HaloFit}
\citep[e.g.][]{halofit}. \\

The emulator can also optionally return an estimate of the variance for its
predictions (See Sec.~II.A of \citealt{george} for the exact definition). This can
be useful to estimate the confidence of the emulator in a given part of the
parameter space. As we will show below, this variance is overestimated for low
values of $k$ as we did not attempt to force the emulator to always return $R=1$
at values of $k$ close to our largest scale bins. \\

Our emulator is distributed publicly in the form of a \techjargon{python}
package named \codename{FlamingoBaryonResponseEmulator} and will be made
available on the \flamingo project
web-page\footnote{\url{https://flamingo.strw.leidenuniv.nl/}} or directly via
the \techjargon{PyPi} package index upon acceptance of this manuscript for
publication. Comprehensive documentation and usage examples are provided on the
web-page.

\subsection{Verification and accuracy}
\label{ssec:validation}

Before exploring the results, we start by assessing the quality of the
emulator's predictions for the data points it was trained on.

\begin{figure}
\includegraphics[width=\columnwidth]{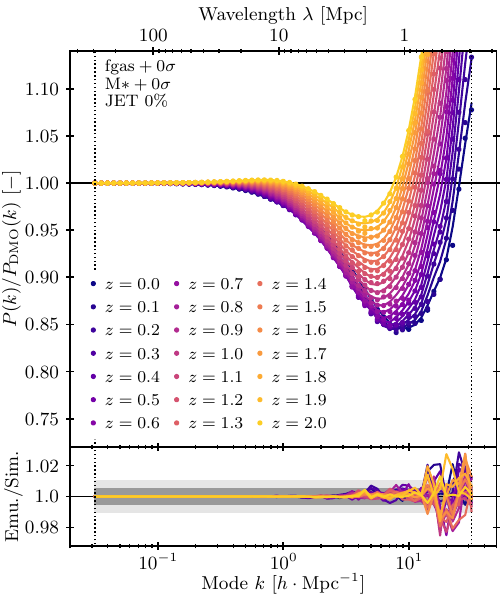}
\vspace{-0.5cm}
\caption{The accuracy of the baryonic response emulator for the \flamingo
  fiducial model (``L1\_m9'') as a function of redshift. The coloured dots show
  the raw power spectra ratios measured directly from the simulations at 21
  different redshifts. The lines show the emulator predictions for the
  corresponding redshifts. Note that the emulator was only trained on data at an
  interval $\Delta z=0.5$. The vertical dotted lines indicate the $k$-range over
  which the emulator was trained. The bottom panel shows the ratio between the
  emulator prediction and the raw simulation output. The shaded regions
  correspond to fractional errors of $0.5$ and $1$ per cent respectively. For all
  redshifts and for all $k < 10~h \cdot {\rm Mpc}^{-1}$, the emulator is
  accurate to better than $1$ per cent.}
\label{fig:verification:fiducial}
\vspace{-0.3cm}
\end{figure}

In Fig.~\ref{fig:verification:fiducial}, we show the baryon response for our
fiducial model (fgas$+0\sigma$, $M_*+0\sigma$, jet $0\%$) at $21$ different
redshifts and for $61$ different values of $k$. Recall that we only trained on
$7$ redshifts and $31$ bins in $k$. For each redshift, the dots indicate the raw
results from the simulation whilst the lines show the predictions of the
emulator. In the bottom panel, we present the ratio between the raw simulation
results and the emulator predictions, with the grey shaded regions indicating
$0.5$ and $1$ per cent differences respectively. 

Before discussing the performance of the emulator, we analyse the evolution with
redshift of the response. We find that the response becomes stronger as the
simulation evolves. The position of the minimum of the response moves from
$k=6~h\cdot{\rm Mpc}^{-1}$ at $z=2$ to $k = 10~h\cdot{\rm Mpc}^{-1}$ at
$z=0$. The range of scales affected by baryons also increases with decreasing
redshift. At $z=2$, the response is negligible up to scales $k \approx
1~h\cdot{\rm Mpc}^{-1}$, with the departure from unity (i.e. no response)
shifting to $k \approx 0.1~h\cdot{\rm Mpc}^{-1}$ by $z=0$.

Turning now to the verification of our model, we find that for all redshifts
shown and for all $k < 10~h\cdot{\rm Mpc}^{-1}$, the emulator described above can
reproduce the simulation data with an accuracy better than $1$ per cent. At larger
values of $k$, we find that the emulator is less accurate but still returns an
answer with a relative error better than $2$ per cent.

\begin{figure}
\includegraphics[width=\columnwidth]{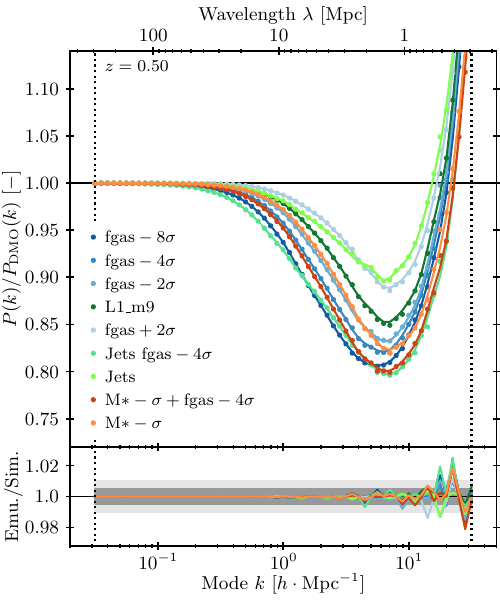}
\vspace{-0.5cm}
\caption{The accuracy of the baryonic response emulator for the \flamingo
  feedback variations (the runs listed in Table~\ref{tab:run_variations} with
  the line colours matching the convention of \citealt{Schaye2023}) it was trained
  on at $z=0.5$. The coloured dots show the raw power spectra ratios
  measured directly from the simulations. The lines show the emulator predictions for the
  corresponding model. The vertical dotted lines indicate the $k$-range over which
  the emulator was trained. The bottom panel shows the ratio between the
  emulator prediction and the raw simulation output. The shaded regions
  correspond to fractional errors of $0.5$ and $1$ per cent respectively. For all
  the models and for all $k < 10~h \cdot {\rm Mpc}^{-1}$, the emulator is
  accurate to better than $1$ per cent.}
\label{fig:verification:variations}
\vspace{-0.3cm}
\end{figure}

In Fig.~\ref{fig:verification:variations}, we show the performance of the
emulator for the other models it was trained on. We arbitrarily picked
$z=0.5$ for the figure but verified that the results are identical at other
redshifts. The conclusion here is similar to the one for the previous exercise:
the emulator reaches a relative accuracy better than $1$ per cent for $k <
10~h\cdot{\rm Mpc}^{-1}$ and degrades to around $2$ per cent on smaller scales.

To be more quantitative, we measured the relative difference between the
emulator predictions and the raw simulation results for all $z\leq3$ at which we
have data and for all nine feedback models. We recorded the maximal error
reached and found that at $k < 10~h\cdot{\rm Mpc}^{-1}$, this never exceeds
$1.1$ per cent. At $k < 3~h\cdot{\rm Mpc}$ and for $0 \leq z \leq 3$, the most
relevant range for current cosmology measurements, the relative error of the
emulator is always smaller than $0.25$ per cent.

We can thus conclude that the emulator designed in the previous section
faithfully reproduces the raw simulation results. Unless stated
otherwise, the rest of this paper will only use the emulator to show
results and not the raw simulation data.

\subsection{Alternative description of input parameters}
\label{ssec:alternative}

As discussed in Sec.~\ref{ssec:flamingo} and more thoroughly in \cite{Kugel2023},
the \flamingo simulations were calibrated to match the inferred gas fractions
from a combined data set of X-ray and weak-lensing. We then uniformly shifted
the data by a certain number of $\sigma$ and calibrated the model to match these
shifted data sets. The ``fgas$\pm N\sigma$'' input parameter of our emulator
corresponds to the $\sigma$-shift with respect to the original data set used by
\cite{Kugel2023}. Instead of referring to models defined with respect to a fixed
data set, it may be advantageous to instead use the absolute gas fractions in
groups or clusters as the input parameter of the emulator.

\cite{VD2020} showed that across many galaxy formation models, a tight
relationship exists between the \emph{baryon} fraction in a certain halo mass
bin and the baryonic response at a fixed scale $k$. Different halo masses then
dominate the effect at different $k$ values \citep[see
  also][]{vanLoon2024}. Using these fractions is thus a meaningful input
parameter that can also be directly related to observables. Similarly, the
response emulator from the \simulationname{Antilles} suite \citep{Salcido2023}
also uses the baryon fraction (either at a fixed halo mass or across a range of
masses) as input to their model. However, given the limited range of models
varying the stellar fractions in the \flamingo suite, we choose to instead use
the \emph{gas fractions} as a the parameter we vary and leave the exploration of
more general variations where we alter both the stellar and gas factions (and
thus the baryon fractions) to a future study.

\begin{figure}
\includegraphics[width=\columnwidth]{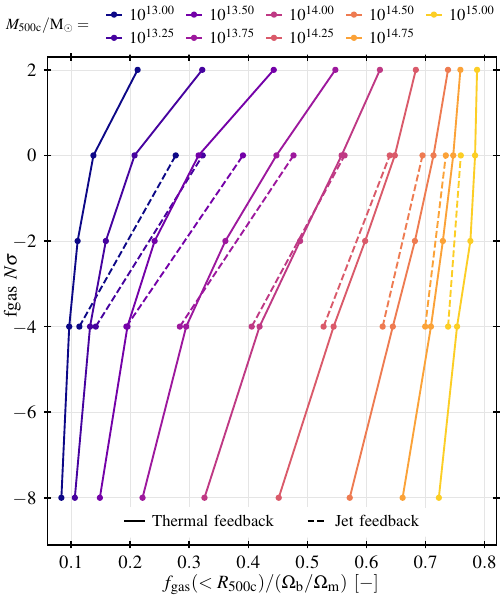}
\vspace{-0.5cm}
\caption{The value of the ``fgas$\pm N\sigma$'' input parameter to our emulator as a
  function of the $z=0$ gas fraction within $R_{500{\rm c}}$ of the simulated
  clusters for different halo masses (different line colours). The solid lines
  correspond to the values extracted from the simulations with the thermal AGN
  model whilst the dashed lines show the values for the simulations exploiting
  the jet-based AGN model.}
\label{fig:alternative:gas_fractions}
\vspace{-0.3cm}
\end{figure}

As various observational data sets are able to constrain the gas fractions in
different halo masses and because the \flamingo simulations may not necessarily
match the trend with mass of all these observations, we decided not to
pre-define a specific halo mass for which the gas fraction is used as input to
our model.  We, instead, choose to present the connection between the gas
fraction at $z=0$ and the actual input parameter ``fgas$\pm N\sigma$'' in
Fig.~\ref{fig:alternative:gas_fractions} and let users of our package decide
based on their available observational input data which halo mass best suits
their needs.

In Fig.~\ref{fig:alternative:gas_fractions}, we show, using different line
colours for each halo mass, the mapping between the gas fractions in the
simulation at that halo mass and the input parameter ``fgas$\pm N\sigma$'' of our
emulator. We use $M_{500{\rm c}}$ as our halo mass definition and report the
gas fractions (normalised by the cosmic mean) within the corresponding
over-density radius $R_{500{\rm c}}$. This choice was made to match the radii
commonly used by cluster studies.  The dashed lines correspond to the
simulations using the jet AGN implementation. Note that since this model was
only run for two different values of ``fgas$\pm N\sigma$'', the mapping at more
extreme values is not available.

For intermediate masses, $M_{500{\rm c}}\sim10^{14}{\rm M}_\odot$, the two
models for AGN feedback agree (i.e. the solid and dashed lines
overlap). This is largely a consequence of the calibration effort as this is the
mass range that was the most constraining in the data set used by
\cite{Kugel2023}. At higher, and especially at lower masses, the two models
start to differ. This, in turn, leads to differences in the baryonic response
these models generate for a fixed ``fgas$\pm N\sigma$'', as can be seen in
Fig.~\ref{fig:verification:variations} when comparing the ``Jets'' and
``L1\_m9'' models (both using ``fgas$+0\sigma$'') or the two models calibrated
to ``fgas$-4\sigma$''. \\

One can then use Fig.~\ref{fig:alternative:gas_fractions} and the gas fractions
obtained from observed data sets to set a prior on the range of values our
parameter ``fgas$\pm N\sigma$'' can take; for instance when attempting to
marginalise over baryon effects for cosmology inference.

\section{Results}
\label{sec:results}

Having demonstrated that the emulator we built is sufficiently accurate, we now
turn to the exploration of some of the baryonic response predictions of the
\flamingo model.

\subsection{The baryonic response in \flamingo}
\label{ssec:baryonic_response}

In this section, we present a selection of results at $z=0$ by varying the
emulator's input parameters one by one.

\subsubsection{Varying the gas fraction for the thermal AGN model}

\begin{figure}
\includegraphics[width=\columnwidth]{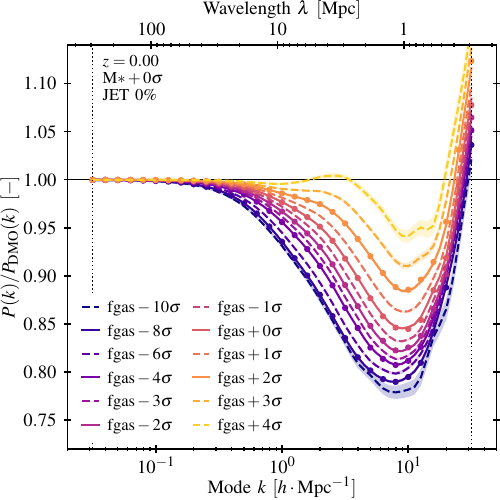}
\vspace{-0.5cm}
\caption{Predictions of the \flamingo baryon response emulator at redshift zero
  for different models using thermal AGN feedback deviating from the X-ray
  inferred gas fractions the simulations were designed to reproduce and
  expressed as the number of sigma discrepancy between the calibrated gas
  fraction and the data. The variables kept fixed in the emulator are displayed
  on the top left. The dots correspond to raw data from the simulations with the
  solid lines showing the emulator prediction at the same value of ``${\rm
    fgas}\pm N\sigma$''. The dashed lines show the emulator predictions for
  interpolation or extrapolation beyond the simulations used for its
  construction. The shaded regions show the 2-sigma uncertainty of the emulator
  prediction.}
\label{fig:results:thermal_variations}
\vspace{-0.3cm}
\end{figure}

We start by studying the effect of changing the ``fgas$\pm N\sigma$'' parameter
for the purely thermal AGN model while keeping the stellar masses at their
fiducial values. In Fig.~\ref{fig:results:thermal_variations}, we vary this
parameter from $-10$ to $+4$, i.e. going from a model that would lead to gas
fractions in clusters $10\sigma$ below the data used for calibration to a model
$4\sigma$ above it. The different line colours correspond to the responses for
the different parameter values. We use solid lines to indicate models for which
we have a simulation and which were thus part of the training set. For these, we
additionally show using dots the actual simulation data. The dashed lines
correspond to the emulator's predictions for parameter combinations where no
simulation currently exist. They are thus interpolations or extrapolations
beyond the training set. Finally, we indicate using a shaded region the
$2\sigma$ error estimate on the prediction reported by the emulator \citep[for
  a formal definition see][]{george}.

As can be seen, the response becomes larger as the value of the parameter
``fgas$\pm N\sigma$'' decreases. This is, in agreement with previous studies
linking the baryon content of clusters to the intensity of the response in
simulations \citep[e.g.][]{Semboloni2011,Semboloni2013,VD2020, Salcido2023} or
from empirical/phenomenological models \citep[e.g.][]{Schneider2019,
  Debackere2020, Arico2021, Mead2021}. At values of ``fgas$\pm N\sigma$'' below
$-8$, the change in the response becomes smaller and the predicted response
overall reaches a saturation value; meaning the \flamingo model cannot produce
an even stronger response when only this parameter is changed. We note, however,
that we have not run a simulation at that point; this is an extrapolation from
the emulator only.

Similarly, at values of the parameter much above the training regime, we find
that the predicted response from the emulator flirts with the unity line, in a
likely unphysical manner. Whilst understanding the behaviour of the model in
this regime could lead to valuable physical insights, we choose not to spend
more time in this area as the match to observations and cosmological data sets
seems to require a response that is typically stronger than our fiducial model
\citep[e.g.][]{McCarthy2024}, let alone models overshooting the gas fraction
data we calibrated to.

Note also the position of the minimum of the response. For the values of
``fgas$\pm N\sigma$'' explored here, we find that the minimum stays in the range
$k=8$ to $10~h\cdot{\rm Mpc}^{-1}$ with a weak dependence on ``fgas$\pm
N\sigma$''. To first order, the input parameter changes the normalisation of the
response without changing the affected range of $k$. It is instructive to
compare this to Fig.~\ref{fig:verification:fiducial} where we see that the
position of the minimum does decrease systematically with redshift.

\subsubsection{Varying the gas fraction for the jet AGN model}

\begin{figure}
\includegraphics[width=\columnwidth]{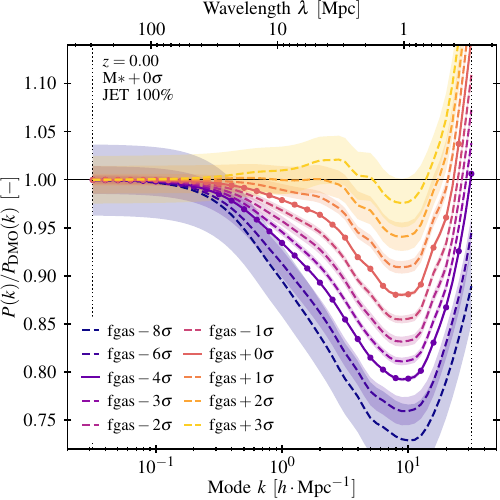}
\vspace{-0.5cm}
\caption{Same as Fig.~\ref{fig:results:thermal_variations} but for the model
  using the AGN jet model. For ``${\rm fgas}\pm N\sigma$'' values beyond the
  training range, the emulator returns a large uncertainty, even at low values
  of $k$. The prediction error is there largely overestimated, mainly due to the
  fact that there are only two simulations sampling this part of the parameter
  space. At fixed value of ``${\rm fgas}\pm N\sigma$'', the emulator predicts a
  larger response than in the case with purely thermal AGN.}
\label{fig:results:jet_variations}
\vspace{-0.3cm}
\end{figure}

We now repeat the exercise of varying ``fgas$\pm N\sigma$'' but for the model with
``100\% jet''. In terms of the 3D parameter space of inputs to our emulator, we
are exploring the same directional vector as above but starting from a position
shifted along the jet fraction axis. The result of this exercise is displayed in
Fig.~\ref{fig:results:jet_variations} where we used the same line style
convention as in the previous figure.

As can be seen, the size of the shaded regions indicating the emulator
uncertainty is significantly larger than for the thermal AGN model (i.e.
$0\%$ jet). This stems from the fact that we have only two runs (the two solid
lines in the figure) using the jet AGN model and the predictions made in this
area of the parameter space are thus more uncertain, especially when
extrapolating beyond the training range. Note also that we do not show the
``fgas$-10\sigma$'' and ``fgas$+4\sigma$'' predictions as their large
uncertainties make them lose any significance. The uncertainty at low $k$ is
strongly overestimated as all physical models have to return to unity in that
regime; at least for the range of parameters explored here. This large
uncertainty is a consequence of keeping our emulator simple with no physical
insights that would allow it to return a more physically-motivated uncertainty
in the low-$k$ regime. It is possible that a different choice of mean model for
the emulator would have led to a smaller error estimate. We do, however, prefer
to keep the model as shown here; the level of uncertainty that is acceptable
will be application-dependent. We thus leave the choice of what to use to users
of the emulator.

Overall, the baryonic response obtained for our model with $100\%$ jet shows a
stronger dependence on the input parameter value than for the equivalent model
with $0\%$ jet. For negative values of ``fgas$N\sigma$'', the response is stronger
and for positive values it is weaker than in the thermal AGN case. However, the
$k$-range affected and the position of the minimum are similar.

\subsubsection{Interpolating between AGN models}

\begin{figure}
\includegraphics[width=\columnwidth]{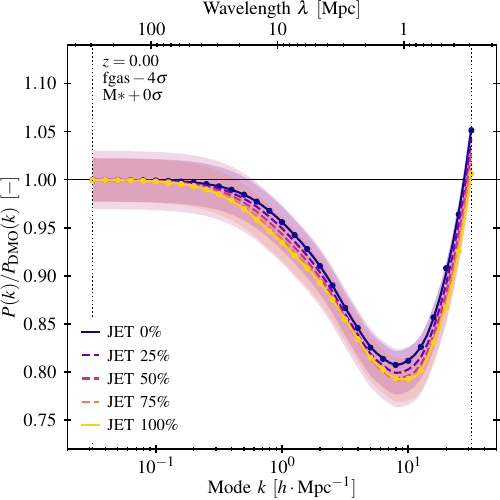}
\vspace{-0.5cm}
\caption{Same as Fig.~\ref{fig:results:thermal_variations} but for predictions of
  the \flamingo baryon response emulator when the fraction of AGN work done by
  the jet model is varied between $0\%$ and $100\%$, here for models fitting the
  gas fraction data shifted by $-4\sigma$. Note that simulations have only been
  performed at the two extremes of the jet fraction range. }
\label{fig:results:jet_fraction_variations_4sigma}
\vspace{-0.3cm}
\end{figure}

The impact of the choice of AGN model is clear from the previous two figures. As
we constructed our emulator with the formally binary jet model
option\footnote{Simulations only exist with the jet model or without it.
No simulation with a fractional jet mechanism exist.} replaced by a continuous
fraction, we can explore the effect of changing the model. We show this in
Fig.~\ref{fig:results:jet_fraction_variations_4sigma}, where we vary the jet
fraction for the case of ``fgas$-4\sigma$''. We picked this value of the gas
fractions for the figure as it shows a larger difference between models than for
the fiducial ``fgas$+0\sigma$'' but the effect is similar at other values. We
note that there is no simple way to generate an actual \flamingo simulation that
has a jet fraction other than $0\%$ or $100\%$. The results shown here are
purely an interpolation constructed from the emulation.

As can be seen from the figure, the response is stronger for a larger jet
fraction. This is consistent with the gas fractions shown in
Fig.~\ref{fig:alternative:gas_fractions}, with the halo masses, at the relevant
$k$-scale, most responsible for the response displaying a lower gas fraction in
the jet model (dashed lines). The uncertainty reported by the emulator is here
again quite large, largely because only two actual simulations span this
dimension of the training set. The uncertainty at low-$k$ is also likely
overestimated given that all models are expected to return to unity there.

\subsubsection{Varying the stellar fractions in the model}

\begin{figure}
\includegraphics[width=\columnwidth]{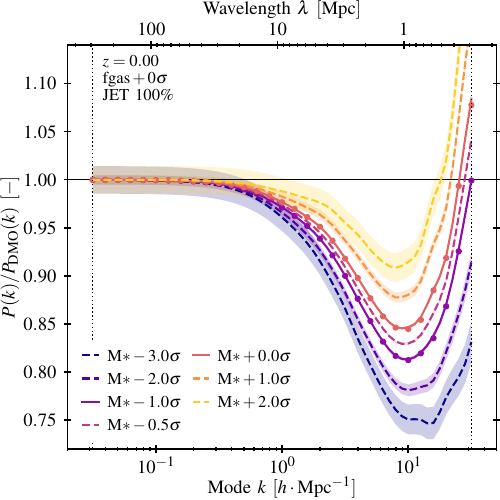}
\vspace{-0.5cm}
\caption{Same as Fig.~\ref{fig:results:thermal_variations} but varying the fit
  to the $z=0$ stellar mass function whilst keeping the gas fraction in clusters
  fixed to the fiducial value.}
\label{fig:results:stellar_fraction_variation}
\end{figure}

Having explored the first two dimensions of the parameter cube, we now turn to
the variation of the last input parameter, the change in the $z=0$ stellar
masses. This corresponds to models where the mass of every galaxy is
lowered/increased by a certain number of $\sigma$ from the original data
\citep[GAMA DR4,][]{Driver2022} that the fiducial \flamingo simulations were
calibrated to. We show the effect of this parameter in
Fig.~\ref{fig:results:stellar_fraction_variation}, where we adopt the same line
style convention as for the last three figures. Along this axis, the number of
actual training points is again very small (two) so the emulator's predictions
become rapidly very uncertain. The uncertainty on the observed data is, however,
much smaller here than for gas fractions and this does leave us with less
freedom to drastically change the amount of stellar mass formed overall in our
simulations. The effect of this parameter is smaller than the changes in the gas
fractions studied above. The uncertainty on the stellar masses in the data only
significantly impacts the baryonic response for $k>1~h\cdot{\rm Mpc}^{-1}$. At
fixed gas fraction in clusters, we find that a lower stellar mass function leads
to stronger baryonic response. This is in line with expectations as it requires
the stars and the AGN to provide more feedback to suppress the formation of
stars in the objects on top of the feedback required to obtain the requested gas
fraction.

As discussed in Sec.~\ref{ssec:alternative}, a future improvement of our emulator
will be to offer more variations of both the gas \emph{and} stellar fractions
independently to have a broader coverage of the possible baryon fraction in
clusters, akin to the model of \cite{Salcido2023}. We leave these improvements
for a future release of more \flamingo simulations extending the parameter space
in these directions.

\subsection{Evolution with redshift}
\label{ssec:redshift_evolution}

In the previous subsection, we focused on the results at $z=0$. As most of the
cosmology surveys use data at higher redshifts and use many bins or bins
spanning a wide redshift range, it is interesting to quantify the evolution of
the response with redshift. Our emulator was trained with data from the
simulations up to $z=3$, a range beyond sufficient for all current and upcoming
surveys. It can thus be directly used in the analysis of survey data.

\begin{figure}
\includegraphics[width=\columnwidth]{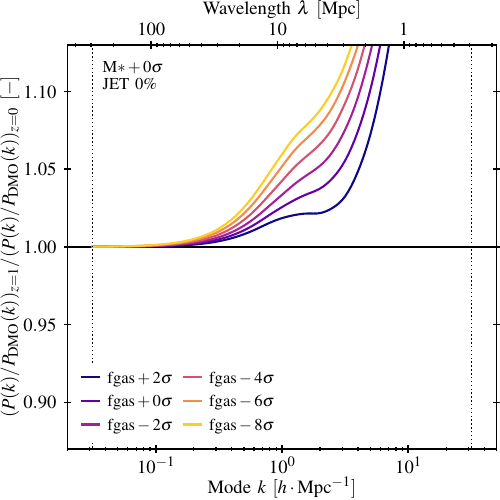}
\vspace{-0.5cm}
\caption{The ratio of the baryon response at redshift $z=1$ and $z=0$ for models
  with different values of the ``${\rm fgas}\pm N\sigma$'' value, as indicated
  in the legend. Note the smaller range of the vertical axis used here. Models
  with no redshift evolution of the response would display a flat line on this
  figure. The models with lower gas fractions in clusters, i.e. with more
  intense feedback, show a stronger evolution with redshift of the predicted
  baryonic response.}
\label{fig:results:redshift_evolution}
\vspace{-0.3cm}
\end{figure}

The evolution of the response, in particular of the $k$-range affected and of
the position of the minimum of the response, with redshift for our fiducial
model was already presented in Fig.~\ref{fig:verification:fiducial}. Besides
this base case, exploring the full range of models discussed in
Sec.~\ref{ssec:baryonic_response} across all redshifts would extend our study
too much, we thus restrict our evolution analysis here to the simple question of
whether models with stronger feedback also display a stronger evolution in
time. To measure this, we take the baryonic response at $z=1$ and compare it to
the response at $z=0$ for different values of the ``fgas$\pm N\sigma$''
parameter. The result of this experiment is displayed in
Fig.~\ref{fig:results:redshift_evolution} where we show the ratio between the
two responses. The different coloured lines correspond to different values of
``fgas$\pm N\sigma$'' (i.e. different feedback strengths). Models displaying no
evolution in their response between $z=1$ and $0$ would be shown as a horizontal
line of value $1$. The further a model is from this line (in either direction),
the more redshift evolution is found. As can be expected, the stronger the
feedback response at $z=0$ (i.e. the smaller the value of ``fgas$\pm
N\sigma$''), the stronger the evolution. The responses for different feedback
strengths thus look more similar at higher redshift. As the response is mainly
driven by the AGN-induced gas flows in large haloes \citep[e.g.][]{VD2011}, we
can qualitatively understand this trend by the need of massive galaxies to first
form and by the time it later takes for the expelled gas to reach large
distances.

We note that we have verified that the same trend holds for models using a jet
fraction $>0$ or a different ``$M_*\pm N\sigma$'' value.

\subsection{Dependence on the cosmological model}
\label{ssec:cosmo_dep}

The emulator was trained on simulations that all assume the same background
cosmology. If the emulator is to be used in cosmology inference using survey
data, this would imply that we \emph{assume} that the response is independent of
the cosmology chosen. Whilst verifying this assumption is a critical task, it is
beyond the scope of this paper and of the current range of simulations present
in the \flamingo suite.

\cite{Elbers2024} explored the correlation between cosmological model and
baryonic response in the \flamingo simulations, especially in the context of
neutrino mass variations. They found a small dependence, which they were able to
explain using the changes in the halo mass-concentration relation induced by the
changes in the cosmology. We do not repeat this exercise here, but we complement
it with an analysis of an additional two simulations using the ``lensing
cosmology'' (LS8) cosmological model ($\Omega_{\rm m}=0.305,~ \Omega_{\rm
  b}=0.0473,~\sigma_8=0.760,~h=0.682,~n_{\rm s}=0.965,~\sum m_\nu c^2= 0.06~{\rm
  eV}$) of \cite{Amon2022}. We ran two simulations with this background
cosmology. The first one using the ``fgas$+0\sigma$'' baryon physics model and
the second one using the ``fgas$-8\sigma$'' one. These two models largely
bracket the responses we obtained in the \flamingo suite. We then compare the
baryonic responses predicted by these simulations to the predictions of the
emulator that was trained on the fiducial cosmology\footnote{Note that the
response for the raw LS8 simulations was obtained by taking the ratio of the
matter power-spectrum with respect to the matter power spectrum in a DMO
simulation using the same cosmology.}.

\begin{figure}
\includegraphics[width=\columnwidth]{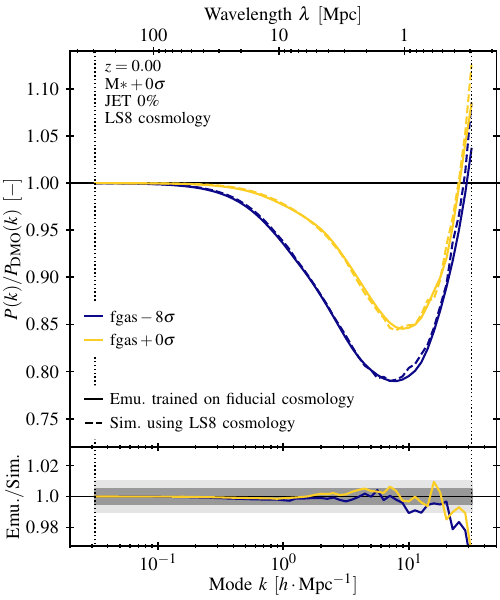}
\vspace{-0.5cm}
\caption{Comparison of the matter power spectrum baryonic response from our
  emulator (solid line) trained on our runs using our fiducial cosmology to the
  responses obtained from simulations using our low $S_8$ (LS8) cosmology
  (dashed lines). The bottom panel shows the ratio of the prediction of the
  emulator, trained on our fiducial cosmology model only, to the two simulations
  we have run using the LS8 cosmology. The shaded regions indicate $<0.5$ and
  $<1$ per cent relative difference with respect to the raw simulation
  prediction. For the relevant range of $k$-scales and for the two models which
  span a wide range of possible responses within the \flamingo suite, we find
  that the cosmology dependence of the baryonic response is below $1$ per cent
  of the total signal.}
\label{fig:results:cosmology_variations}
\vspace{-0.3cm}
\end{figure}

We show this comparison in Fig.~\ref{fig:results:cosmology_variations}. The
emulator predictions (trained only on our fiducial cosmology) are shown as solid
lines whilst the raw simulation data is displayed using dashed lines. The bottom
panel shows the ratio between the two to highlight the differences. For both
feedback models (different colours), we find that the relative difference
between the emulator predictions (trained on our fiducial cosmology) and the raw
simulation data is smaller than $1$ per cent for all $k < 10~h\cdot{\rm
  Mpc}^{-1}$. This indicates that the baryon response in \flamingo is only
slightly affected by the choice of cosmological model. The figure shows this
comparison at $z=0$ but we verified that the same conclusion holds also at
higher $z$.

Interpreting this result in the light of the analytic model of
\cite{Elbers2024}, we can understand the small difference due to cosmology as
coming from the small change in the mass-concentration relation of haloes
between our fiducial cosmology and the LS8 one \citep[as expected from models of
  the concentration such as][]{Correa2015}. From the analysis of
\cite{Elbers2024}, we do expect to find larger differences (up to a few percent
at $k=10~h\cdot{\rm Mpc}^{-1}$) for variations within $\Lambda$CDM that affect
the mass-concentration relation more (see their Sec.~4). Note that one would
also expect a dependency of the response on the ratio $\Omega_{\rm
  b}/\Omega_{\rm m}$. Large variations in this ratio are not present in the
current set of \flamingo simulations.\\

We leave the full exploration of the cosmological dependence of the power
spectrum response to a future study, where we will expand the emulator to also
predict the response as a function of cosmological parameters.

\subsection{Comparison to other studies}
\label{ssec:comparison_others}

We conclude our description of the baryonic response in \flamingo with a
comparison of our model to results from other simulations and models. We
restrict our analysis to $z=0$ but understanding the evolution of the response
with time will be important for current and future surveys where data from a
wide range of redshifts enters the analysis.

\subsubsection{Comparison to hydrodynamical simulations}

\begin{figure}
\includegraphics[width=\columnwidth]{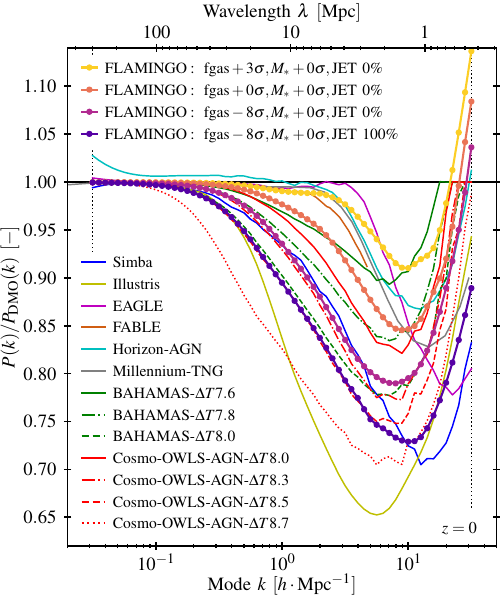}
\vspace{-0.5cm}
\caption{Comparison of the $z=0$ matter power spectrum baryonic response in four
  \flamingo models (from our emulator) to the response obtained in a selection
  of simulations from the literature (see text for details). Depending on the
  input parameters, the response returned by the \flamingo emulator can be
  similar to many published models; only the models with the strongest feedback
  are beyond the range of our emulator. Of particular interest are the two
  models at ``fgas$-8\sigma$'' using the thermal and jet models which closely
  resemble the \cowl models with $\Delta T_{\rm AGN}=8.3$ and $8.5$ respectively
  that are often used in the literature as representatives of the baryonic
  response in hydrodynamical simulations. }
\label{fig:results:comparisons_others}
\vspace{-0.3cm}
\end{figure}

In Fig.~\ref{fig:results:comparisons_others} we compare four models extracted
from our emulator to a selection of simulations from the literature. We use here
the compilation of homogenised data from \cite{VD2020}, expanded to more recent
runs. In particular, we show the results from \simulationname{Simba}
\citep{Dave2019}, \simulationname{Illustris} \citep{Vogelsberger2014},
\simulationname{Eagle} \citep{Schaye2015}, \simulationname{Fable}
\citep{Henden2018}, \simulationname{Horizon-AGN} \citep{Chisari2018},
\simulationname{Millennium-TNG} \citep{Pakmor2023}, three variations of
\simulationname{BAHAMAS} \citep{BAHAMAS}, and four variations of
\simulationname{Cosmo-OWLS} \citep{LeBrun2014}. Note that, as discussed in
Sec.~\ref{ssec:convergence}, some of these simulations exploited volumes that are
likely too small to produce converged results.

As can be seen, the range of predictions from all these models is relatively
broad, as is the baryon fraction in clusters they predict
\citep{VD2020}. Nevertheless, our emulator can cover a large range of models
with its predictions. Only the two most extreme models
(\simulationname{Illustris} and \simulationname{Cosmo-OWLS-AGN-$\Delta T8.7$})
are beyond the reach of our emulator, and likely of what the \flamingo family of
simulations can produce without alterations to the physics.

It is interesting to note that our models with ``fgas$-8\sigma$'' using the
thermal and jet models closely resemble the \cowl models with $\Delta T_{\rm
  AGN}=8.3$ and $8.5$ respectively. The latter of these two models is often used
as a representative of a large baryonic response when analysing data sets
\citep[e.g.][]{Mead2020, Amon2022, Preston2023, Bigwood2024}. Fine tuning of the
parameters of the emulator could likely lead to our model matching other
simulations shown here. We thus conclude that our family of models and the
emulator constructed on top of it are able to cover almost the entire range of
predictions from the literature. Missing the strongest responses is, however,
possibly a limitation as recent studies seem to suggest stronger responses are
required to explain the data \citep{Amon2022, Preston2023}, in particular kSZ
observations \citep{Bigwood2024, Hadzhiyska2024, McCarthy2024} and the tSZ power
spectrum \citep{McCarthy2014, McCarthy2018, McCarthy2023}.

With its connection to gas fractions at specific halo masses
(Fig.~\ref{fig:alternative:gas_fractions}), our emulator can be meaningfully be
related to observables. These can in turn be used to provide a prior on the
range of responses compatible with the data. The more meaningful labelling of the
models via ``fgas$\pm N\sigma$'', as compared to a label based on subgrid
parameter values, helps in this respect.

\subsubsection{Comparison to the \antilles suite}

\begin{figure}
\includegraphics[width=\columnwidth]{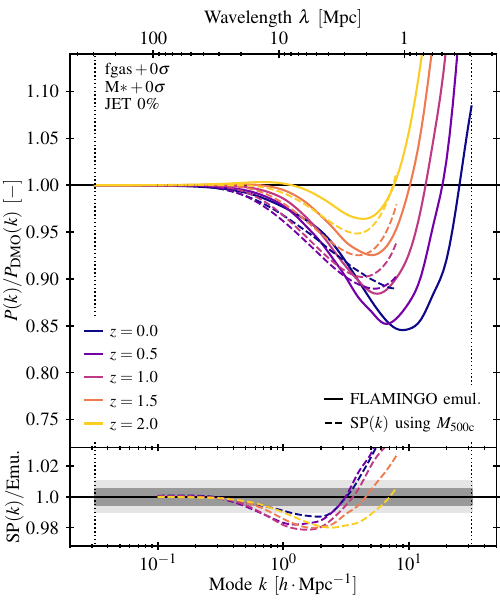}
\vspace{-0.5cm}
\caption{Comparison of the matter power spectrum baryonic response at different
  redshifts in the fiducial \flamingo simulation from our emulator (solid lines)
  to the one predicted by the ${\rm SP}(k)$ model of \citet{Salcido2023} fitted
  to the true $M_{500{\rm c}}$ - baryon fraction relation extracted from the
  \flamingo halo catalogs (dashed lines). The bottom panel shows the ratio
  between the ${\rm SP}(k)$ model and our emulator.}
\label{fig:results:comparisons_SPK}
\vspace{-0.3cm}
\end{figure}

Using the \antilles suite of 400 simulations run with the \owl model,
\cite{Salcido2023} constructed a model, ``SP(k)'', predicting the baryonic
response using the baryon fraction in groups and clusters as input parameter to
their model. Their much wider range of simulations allows them to encompass a
much wider range of possible responses than we can with our \flamingo-based
emulator. They offer multiple versions ranging in complexity of the input going
from a single parameter to providing the fractions over a range of halo
masses. We use this latter version here to compare to our models. Specifically, we
measured the baryon fractions as a function of halo mass (note the difference
with the gas fractions we used in Fig.~\ref{fig:alternative:gas_fractions}) for
our fiducial \flamingo model and used them as an input to their model. The
resulting response is shown for five different redshifts in
Fig.~\ref{fig:results:comparisons_SPK} alongside the response obtained from our
emulator.

\begin{figure*}
\includegraphics[width=\textwidth]{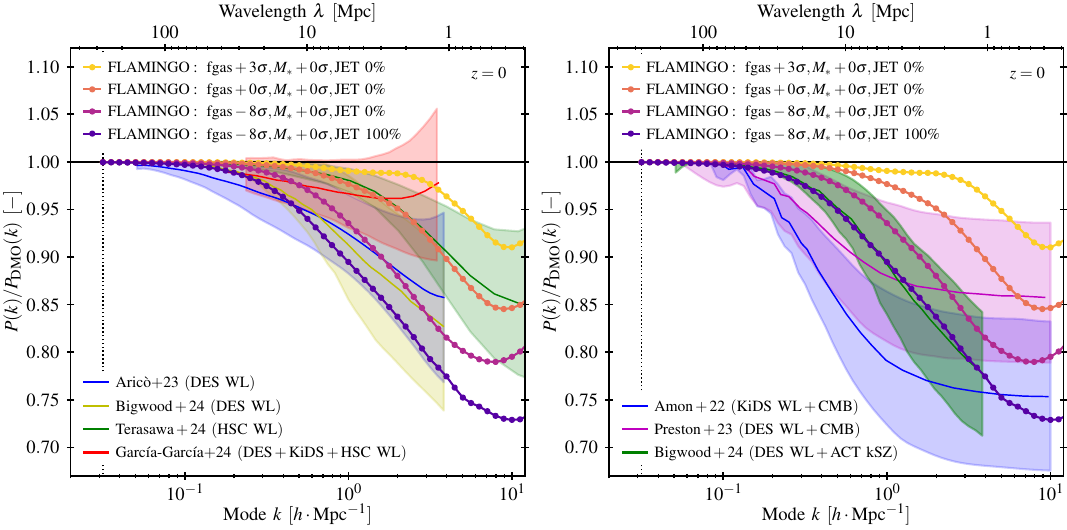}
\vspace{-0.5cm}
\caption{Comparison of the $z=0$ matter power spectrum baryonic response in four
  \flamingo models (from our emulator) to the response inferred from the
  analyses of different weak-lensing cosmology probes combined with a halo model
  or a baryonification model. The right panel shows analyses where in addition
  either primary CMB anisotropy data or kSZ data was used to infer the response
  (see text for details). Note the reduced $k$-range plotted here compared to
  the previous figures.}
\label{fig:results:comparisons_S8}
\vspace{-0.3cm}
\end{figure*}

As can be seen, despite using the exact baryon fractions of the simulations as
input, the SP(k) model predictions do not match the \flamingo results. They
predict a stronger response at $k\approx1~h\cdot{\rm Mpc}^{-1}$ and a weaker one
at $k>3~h\cdot{\rm Mpc}^{-1}$, in particular at low redshift. This indicates
that knowing the baryon fraction in haloes is not a sufficient condition to
obtain the baryonic response. Indeed, \cite{Debackere2020} used a halo model to
show that the baryon fraction beyond $R_{500{\rm c}}$ is important.

The \flamingo and \antilles simulations do not have the same stellar fraction in
their haloes and this likely explains part of the difference here. Note that it
may be possible to explore our emulator to match SP(k) more closely. One could,
for instance, lower ``$M_*\pm N\sigma$'' whilst increasing ``fgas$\pm N\sigma$''
in order to maintain the same baryon fraction in haloes. However, as we do not
have predictions for the stellar fractions as a function of these parameters, we
leave this exercise for a future study with a more comprehensive emulator. This
simple comparison nevertheless indicates that predicting the baryonic response
using a single number (e.g. the baryon or gas fraction) per halo mass bin may
not be sufficient and more complex models will have to be constructed in the
future.

\subsubsection{Can \flamingo feedback solve the $S_8$ tension?}

Over the last few years, precision tests of the \lcdm models, particularly those
exploiting low-redshift weak-lensing and large-scale structure probes, have
reported measurements of $S_8 \equiv \sigma_8\sqrt{\Omega_{\rm m}/0.3}$ in
conflict with the value inferred from the primary CMB anisotropies and BAO
experiments. Depending on the data sets and their analysis, this so-called
tension can reach $3\sigma$ \citep{Heymans2021, Abbott2022, Amon2023,
  Miyatake2023, McCarthy2023, Terasawa2024}.

Among the different solutions put forward to solve this tension, the possibility
that (baryon and galactic feedback) physics on the scales of groups and clusters
contributes to the non-linear effects on the range of scales on which the power
spectrum is probed is an interesting proposition as it does not require an
alteration of the background cosmological model. We explore here whether the
response obtained in the \flamingo model is compatible with the required
baryon-induced non-linear solution to the tension.

\cite{Amon2022} constructed a toy model for the effect of baryons, inspired by
the functional form obtained from the \codename{HMCode2020} \citep{Mead2020},
and constrained it to reconcile the weak-lensing shear analysis of the KiDS-1000
survey and the primary CMB anisotropy data. Using a similar approach,
\cite{Preston2023} combined the DES year 3 shear data to the CMB data to also
constrain the required non-linear power spectrum changes. Both these constraints
are shown as, respectively, the blue and magenta shaded regions in the right
panel of Fig.~\ref{fig:results:comparisons_S8} \citep[see also][for an analysis
  combining DES lensing to ROSAT X-ray all-sky photon counts]{Ferreira2024}. A
parallel approach is to exploit the weak-lensing data alone in combination with
a halo model or emulator for the baryonic response. The analysis of
\cite{Arico2023} and \cite{Bigwood2024}, both exploiting the DES year 3 data but
with different emulators --- \codename{BACCO} \citep{Arico2021}, 
\codename{BCemu} \citep{Giri2021}, respectively --- and different analysis
pipelines, are shown as shaded regions in the left panel of
Fig.~\ref{fig:results:comparisons_S8} alongside the results of
\cite{GarciaGarcia2024} also using the \codename{BACCO} emulator but combining
data from DES, KiDS and HSC. The results of \cite{Terasawa2024} exploiting HSC
data and the \codename{HMcode2020} halo model are also shown. Finally,
additional datasets informing the emulators or models on the gaseous content of
haloes can be used. The constraints of \cite{Bigwood2024} including kSZ data
from ACT are shown in the right panel of Fig.~\ref{fig:results:comparisons_S8}
\citep[see also][]{Schneider2022}. In both panels, we show the response obtained
for four models spanning the range of \flamingo emulator input parameter values.

As can be seen, the more extreme models in the \flamingo suite are compatible
with some of the inferred baryonic responses in these studies. In particular,
both ``fgas$-8\sigma$'' models (i.e. with thermal AGN and with
collimated jet AGN) are within the ranges demanded by the analysis of
\citealt{Bigwood2024} (with or without added kSZ constraints). Interestingly, these
\flamingo models display gas fractions that are in strong tension with the gas
fractions in groups and clusters inferred from joint X-ray and weak-lensing
analyses \citep{Kugel2023}. Putting aside the option of systematic error or
selection effects affecting either data sets, this could indicate that more
complex physics, for instance more advanced feedback models, are required in
simulations to jointly match the X-ray and kSZ measurements \citep[see
  also][]{Hadzhiyska2024,McCarthy2024}. On the other hand, the constraints of
\cite{Terasawa2024} and \cite{GarciaGarcia2024}, both inferring mild baryonic
effects, are well within the range of models available within the \flamingo
suite.

The constraints on the power spectrum response derived by \cite{Amon2022} and to
a lesser extent \cite{Preston2023}, as well as the range $k\lesssim0.5~h\cdot{\rm
  Mpc}^{-1}$ of the constraints from \cite{Arico2023}, are out of reach of our
emulator and likely of what can be achieved using the \flamingo model without
altering the sub-grid models. If the Universe does look like a \flamingo model
with extreme feedback (e.g. ``fgas$-8\sigma$''), the tension between the CMB and
the low-redshift small-scale probes would nevertheless be reduced, though a more
quantitative analysis of this change is left for future studies. Note, however,
that the $S_8$ tension also manifests itself in other probes not related to
weak-lensing, such as in the thermal SZ power spectrum \citep[see
  e.g.][]{McCarthy2014, McCarthy2018, McCarthy2023} and that altering the
small-scale matter power spectrum may not be sufficient.

\section{Conclusions}
\label{sec:conclusions}

In this study, we used the hydrodynamical simulations from the \flamingo project
\citep{Schaye2023, Kugel2023} and extracted their total matter power spectra,
which were then compared to ones taken from the dark matter only counterparts of
the simulations. By taking the ratio of the two matter power spectra, we
obtained the baryonic response of the matter power spectrum generated by
baryonic physics associated with galaxy formation. We then constructed a
Gaussian process emulator using four parameters: the redshift and the three
parameters describing baryon physics of the simulations as input parameters to
reproduce the simulated baryonic response in the range of scales $k=10^{-1.5} -
10^{1.5}~h\cdot\rm{Mpc}^{-1}$ and redshift range $z=0 - 3$, sufficient for all
current and planned large-scale structure probes. These parameters are the
offset of the gas fractions in clusters with respect to the simulation
calibration data (``fgas$\pm N\sigma$''), the offset of the galaxy masses used
for the calibration (``$M_*\pm N\sigma$'') and the fraction of AGN feedback
output in the form of collimated jets. We then explored the accuracy of the
emulator thus constructed and the general predictions for the baryonic response
in the \flamingo simulations. Our findings can be summarised as follows:

\begin{itemize}
  \item The baryonic response at $z=0$ and $z=1$ is converged to better than $1$
    per cent only for simulations using volumes in excess of
    $200^3~\rm{Mpc}^3$. A convergence to better than $0.5$ per cent is only
    achieved for volumes larger than $400^3~\rm{Mpc}^3$ and the higher redshift
    results display a difference with respect to the converged answer on larger
    $k$-scales than at lower redshift (Fig.~\ref{fig:simulations:convergence}).

  \item The emulator constructed from nine \flamingo models is able to reproduce
    the results of the simulations with a relative error of less than $1$
    per cent for scales up to $k = 10~h\cdot\rm{Mpc}^{-1}$ and for all redshifts
    up to $z=3$ (Figs.~\ref{fig:verification:fiducial} \&
    \ref{fig:verification:variations}).

  \item The emulator requires less than $1~{\rm ms}$ per invocation on a single
    compute core to provide the response at all $k$ values. It can thus be used
    as part of a model inference exercise for surveys without penalty.

  \item The ``fgas$\pm N\sigma$'' input parameter to our model can be related to
    the gas fraction in groups/clusters and this latter quantity can be
    extracted from observations to serve as prior on the range of inputs to our
    emulator. The mapping between the two quantities for different halo masses
    is given in Fig.~\ref{fig:alternative:gas_fractions}.

  \item We find the baryonic response to be stronger for models with lower gas
    fractions in groups and clusters (parameter ``fgas$\pm N\sigma$'',
    Fig.~\ref{fig:results:thermal_variations} \&
    \ref{fig:results:jet_variations}) and for lower stellar fractions
    (Fig.~\ref{fig:results:stellar_fraction_variation}).

  \item The response is stronger, at fixed gas fractions in clusters, for the
    models using collimated AGN over thermal AGN feedback
    (Fig.~\ref{fig:results:jet_fraction_variations_4sigma}).

  \item The baryonic response is stronger at lower redshift with the
    range of $k$-scales affected growing as the redshift
    decreases. Simultaneously, the position of the minimum of the
    response moves to larger $k$ (Fig.~\ref{fig:verification:fiducial}).
    
  \item Models with stronger feedback (and thus a stronger response) display a
    larger evolution in their response between $z=1$ and $z=0$
    (Fig.~\ref{fig:results:redshift_evolution}).
    
  \item The dependence of the response on the cosmological model is small ($<1$
    per cent for all relevant $k$, Fig.~\ref{fig:results:cosmology_variations})
    but we only probed two specific cosmological models, leaving room for a
    stronger dependence when other parameters are changed.

  \item By varying its input parameters, our emulator can cover a wide range of
    responses found in simulations from the literature
    (Fig.~\ref{fig:results:comparisons_others}) except the most extreme
    models. The connection between our input parameters and the gas fraction in
    haloes also allows for a more physically meaningful description of the
    response, compared with the use of subgrid parameter values.

  \item Comparing our model to the model obtained from the \antilles
    simulations, we find that, even when using the simulated baryon fractions as
    input parameters, the SP(k) model does not match our results
    (Fig.~\ref{fig:results:comparisons_SPK}). This indicates that more than one
    parameter is necessary to describe the baryonic response.

  \item When comparing our emulator to the constraints on the baryonic response
    derived from either the analysis of weak-lensing with halo models,
    baryonification models, or when combined with additional datasets, we find
    that the most extreme \flamingo models can match some of these constraints
    (Fig.~\ref{fig:results:comparisons_S8}). These models are, however, in
    tension with the cluster X-ray data used to constrain our fiducial model.

\end{itemize}
~\\

We plan to extend our emulator in the future to provide more independent
variations in the gas and stellar fractions in clusters as well as including
changes in cosmological parameters. This latter improvement will allow us to
break free from the assumption that the baryonic response is separable from
cosmology.\\

The emulator in its present form should nevertheless be sufficient
for the analysis of current cosmological surveys where the separability
assumption is still commonly made.

\section*{Acknowledgements}

We thank Josh Borrow for merging, packaging, and releasing changes to the
\codename{swiftemulator} package facilitating the construction and use of our
emulator.

\noindent This work used the DiRAC@Durham facility managed by the Institute for
Computational Cosmology on behalf of the STFC DiRAC HPC Facility
(\url{www.dirac.ac.uk}). The equipment was funded by BEIS capital funding via
STFC capital grants ST/K00042X/1, ST/P002293/1, ST/R002371/1 and ST/S002502/1,
Durham University and STFC operations grant ST/R000832/1. DiRAC is part of the
National e-Infrastructure.

\section*{Data Availability}

The model introduced in this paper will be made fully publicly available on the
\flamingo project's
web-page\footnote{\url{https://flamingo.strw.leidenuniv.nl/}} upon publication
of this manuscript. The raw matter power spectra used to construct the emulator
are available on the same website.

The full \flamingo simulation data will eventually be made publicly available,
though we note that the data volume (several petabytes) may prohibit us from
simply placing the raw data on a server. In the meantime, people interested in
using the simulations are encouraged to contact the corresponding author.

\bibliographystyle{mnras}
\bibliography{bibliography} 




\bsp	
\label{lastpage}
\end{document}